# DYNAMIC MODELING AND CONTROL SYSTEM ANALYSIS FOR CONTINUOUS-DISC FILTERS IN PULP MILL OPERATIONS

José M. Campos-Salazar [1], Felipe Santander [2], Sebastián Larraín [3]

[1] *Reliability Department. Celulosa Arauco y Constitución SA. Chile*
[2] *Process Department. Celulosa Arauco y Constitución SA. Chile*
[3] *Process Research Department. Bioforest SA. Chile*

## ABSTRACT

Vacuum disc filtration is critical in pulp mills for white liquor clarification and pulp washing, involving tightly coupled dynamics between rotational speed, vacuum pressure, slurry concentration, filtrate flow, and cake thickness. These nonlinear interactions are often regulated using empirical methods, lacking formal modeling and control.

This article develops a dynamic, multivariable model of a continuous-disc filter (CD-filter) system based on first principles, simplified to a single representative disc for tractability. A linearized state-space model supports the design of two control strategies: a decentralized PI-based scheme and a centralized model predictive control (MPC). MATLAB-Simulink simulations reveal that MPC outperforms PI in tracking accuracy, overshoot reduction, and disturbance rejection. A 3D efficiency surface illustrates the importance of coordinating inlet flow and solids concentration. Results highlight the need for advanced multivariable control in optimizing CD-filter performance.

## 1. INTRODUCTION

Continuous disc filters (CD-filters) are critical to industrial solid-liquid separation, especially in pulp and paper mills, where they are widely used for pulp washing and white liquor clarification [1–3]. These systems consist of multiple porous discs rotating through a slurry, using vacuum pressure to extract liquid and form a fiber cake on the disc surface. As the cake rotates through dewatering and discharge stages, its thickness, dryness, and formation rate depend on variables such as rotational speed, vacuum pressure, and slurry concentration [4,5]. Proper operation of CD-filters directly impacts mill throughput, filtrate clarity, and chemical recovery efficiency.

Despite their widespread use, CD-filters are underrepresented in modeling and control literature [6–9]. Most existing studies are empirical or steady-state in nature, lacking formal dynamic models that capture the nonlinear, coupled behavior of cake formation, filtrate flow, and mechanical dynamics. In industrial settings, filters are often controlled using decentralized PI loops or operator-based adjustments, which are insufficient to manage multivariable interactions and process constraints.

This work addresses that gap by developing a multivariable nonlinear model of the CD-filter process and implementing two control strategies: a conventional PI-based multiloop control, and a centralized model predictive control (MPC) scheme. The MPC exploits a linearized state-space model to anticipate system behavior and coordinate inputs, offering performance gains in speed regulation, cake dryness control, and vacuum stabilization [10–13].

The remainder of the paper is structured as follows: Section 2 describes the CD-filter architecture and key solid-liquid separation variables. Section 3 formulates the coupled nonlinear dynamic model. Section 4 details the design of decentralized PI and centralized MPC controllers from linearized state-space representations. Section 5 presents closed-loop simulation results under dynamic scenarios. Section 6 concludes with performance insights and control implications for industrial optimization.

## 2. PROCESS DESCRIPTION

***Corresponding author:*** *José M. Campos-Salazar. Reliability Department. Celulosa Arauco y Constitución S.A.. Chile. e-mail.jose.campos.s@arauco.com*

A CD-filter is a rotary vacuum filtration device widely used in pulp and paper mills for applications such as pulp washing and white liquor clarification [1,8]. It consists of a horizontal rotating shaft equipped with multiple porous filter discs—typically 15 in industrial configurations—partially immersed in a slurry tank. As each disc rotates, vacuum pressure is applied to internal compartments via a distribution valve, drawing fluid through the filter media while retaining suspended solids on the disc surface, thereby forming a growing filter cake. The incoming suspension—raw white liquor (RWL)—is introduced with a volumetric flow rate $f_{in}(t)$ and solids concentration $C_{in}(t)$, while the clarified output—filtered white liquor (FWL)—is collected through internal channels. A third stream, known as the mud flow, conveys the removed cake material discharged by a scraper or high-pressure wash system [9].

To enable rigorous dynamic modeling and tractable control system design, this work adopts a technical equivalent model in which the entire multi-disc assembly is represented by a single representative disc. This assumption captures the average behavior of the system, while significantly reducing the modeling complexity. Similar modeling simplifications have been validated in the context of filtration units and continuous processes with symmetrical distributed dynamics [3,8,10]. Moreover, this approach aligns with prior methodologies adopted in simulation environments such as those described by [3], where full-scale rotary disc filters are represented through lumped-parameter models.

As illustrated in Figure 1, the system dynamics are governed by five tightly coupled process variables: rotational speed $\omega(t)$, vacuum pressure $P_v(t)$, solids concentration in the vat $C_R(t)$, cake thickness $H(t)$, and filtrate flow rate $q_f(t)$. Filtration is driven by a pressure differential $\Delta P(t) = P_{atm} - P_v(t)$, and dynamically modulated by resistances $R_c(t)$ (filter cake layer), $R_m$ (medium resistance), and $R_{tot}(t)$ (total resistance). A rise in concentration $C_R(t)$ accelerates cake buildup, which in turn increases resistance and modifies both filtrate flow and vacuum pressure. These dynamic and nonlinear interactions are captured through a multivariable differential equation framework, providing a foundational structure for simulation, control synthesis, and performance optimization in industrial pulp mill environments [3,8–10].

## 3. SYSTEM MODELING

The CD-filter process exhibits inherently dynamic and nonlinear behavior due to the simultaneous interaction between mechanical rotation, vacuum-induced filtration, solids transport, and cake formation. To capture the essential dynamics of the system, a multivariable modeling framework is developed, focusing on five coupled state variables: angular velocity of the shaft, internal vacuum pressure, RWL solids concentration in the vat, cake thickness on the disc surface, and FWL flow rate. These variables influence one another through physical coupling mechanisms such as shear-dependent drag, permeability-driven resistance, and pressure-driven flow. The proposed model builds upon established filtration theory and control-oriented formulations to describe the temporal evolution of each state variable.

To enable a consistent mathematical formulation, several assumptions are adopted: (i) the filter cake is considered incompressible, i.e., constant porosity throughout operation; (ii) the system is modeled as a single equivalent disc, representative of the full 15-disc industrial configuration; (iii) solids removal is treated as continuous rather than discretized by mechanical scraping; (iv) filtration area is assumed constant ($A$), even though in reality it evolves with disc submergence and rotation, i.e., $A(t)$; and vat level control—typically implemented in industrial settings—is omitted in this conceptual model for simplicity. These assumptions strike a balance between physical realism and analytical tractability, allowing the resulting system of nonlinear differential equations to support dynamic simulation, control design, and performance optimization under variable operational conditions [3,8–10].

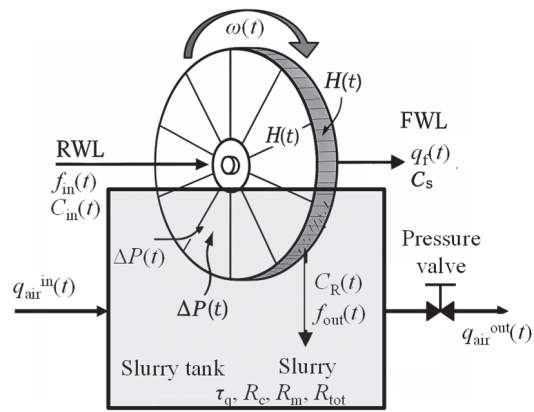

**Figure 1:** CD-filter process. Equivalent diagram.

### 3.1 Rotational Speed Dynamics

The rotational motion of the shaft is driven by motor torque and opposed by the resistive torque resulting from hydrodynamic drag, cake friction, and mechanical losses. The shaft's angular velocity $\omega(t)$ (rad/s) evolves according to:

$$\frac{d\omega(t)}{dt} = \frac{1}{J} \cdot (T_m(t) - T_{res}(t)) \tag{1}$$

Here, $J$ is the total moment of inertia (kg·m²) and $T_m(t)$ is the applied motor torque (N·m). The resistive torque $T_{res}(t)$ increases with higher viscosity, thicker cakes, and faster rotation, reflecting the increased drag as discs rotate through RWL and solid layers [3,8]. This resistance governs the rate of cake formation, since $\omega(t)$ directly affects how long each disc sector remains submerged. A lower $\omega(t)$ allows thicker cakes to form, while a higher speed leads to more frequent cake removal [3]. Also, the $T_{res}(t) = f(\omega(t), H(t))$ is typically modeled as:

$$T_{res}(t) = k_d \cdot \omega(t) + k_c \cdot H(t) \tag{2}$$

where: $k_d$ [N·m·s/rad] is the viscous damping coefficient due to RWL flow and disc hydrodynamics; $k_c$ [N·m/m] is the cake resistance coefficient, related to scraper friction and cake drag, and finally, $H(t)$ [m] is the average cake thickness. Then replacing (2) into (1), gives:

$$\frac{d\omega(t)}{dt} = \frac{1}{J} \cdot (T_m(t) - k_d \cdot \omega(t) - k_c \cdot H(t)) \tag{3}$$

### 3.2 Vacuum Pressure Dynamics

Vacuum pressure, applied on the inner side of the discs, is the primary driving force for filtration [11]. The absolute pressure in the vacuum receiver, $P_v(t)$ (Pa), is governed by the dynamic balance of air ingress and extraction:

$$\frac{dP_v(t)}{dt} = \frac{R_g \cdot T}{V_g} \cdot \left(q_{air}^{in}(t) - q_{air}^{out}(t)\right) \tag{4}$$

where $R_g$ is the universal gas constant (8.314 J/(mol·K)), $T$ is the absolute temperature (K), $V_g$ is the free gas volume in the receiver (m³), $q_{air}^{in}(t)$ is the volumetric inflow of air/vapor from the filter (m³/s), and $q_{air}^{out}(t)$ is the vacuum pressure valve's capacity (m³/s) [11,12]. As filtrate is drawn through the cake and gas permeates the pores, vacuum strength varies with flow resistance. Increased permeability results in higher $q_{air}^{in}(t)$, weakening $P_v(t)$, while enhanced valve performance reduces $P_v(t)$, strengthening the vacuum and increasing FWL flow rate. However, once the cake reaches its maximum permeability, further deepening of the vacuum yields diminishing gains in liquid extraction [2].

### 3.3 Slurry Solids Concentration Dynamics

The dynamic behavior of the suspended solids concentration in the slurry vat (tank) is described by $C_R(t)$ (kg/m³), which denotes the concentration of solids in the slurry vat. The temporal evolution of $C_R(t)$ is governed by a continuous mass balance over the vat volume $V_{vat}$ (m³), expressed as:

$$\frac{dC_R(t)}{dt} = \frac{1}{V_{vat}} \cdot \left(f_{in}(t) \cdot C_{in}(t) - f_{out}(t) \cdot C_R(t)\right) \quad (5)$$

where $f_{in}(t)$ (m³/s) is the volumetric inflow rate of RWL, $C_{in}(t)$ (kg/m³) is the solids concentration in the inflowing slurry, and $f_{out}(t)$ (m³/s) is the slurry flow which includes both filtered liquid $q_f(t)$ and mud withdrawal via cake removal mechanisms.

This balance captures the net accumulation of solids in the vat and reflects the interplay between slurry feed conditions and solids removal processes. Variations in $C_{in}(t)$, filter performance, or mud discharge efficiency directly impact $C_R(t)$, which in turn influences cake formation rates on the disc. External disturbances, such as changes in clarifier performance or upstream carryover of fines [2], can propagate through this dynamic pathway, making $C_R(t)$ a critical variable in filter control and optimization.

## 3.4 Cake Thickness Dynamics

As the discs rotate, solid particles deposit on the filter surface, forming a cake layer of thickness $H(t)$ (m). The dynamics of $H(t)$ arise from the balance between deposition and removal:

$$\frac{dH(t)}{dt} = \frac{1}{\rho_c \cdot A} \cdot C_R(t) \cdot q_f(t) - \omega(t) \cdot H(t) \quad (6)$$

Here, $\rho_c$ is the dry bulk density of the cake (kg/m³), and $A$ is the total filtration area (m²). The first term represents solid deposition, proportional to both the concentration of solids in the slurry vat and FWL flow rate. The second term models cake removal by the scraper, assuming a uniform removal per rotation cycle. Operationally, $\omega(t)$ is adjusted to maintain a desired cake thickness, ensuring effective dewatering without excessive cake buildup [11]. Although the model assumes a constant cake density, compressibility effects can be incorporated for more detailed studies [13].

## 3.5 Filtered White Liquor Flow Dynamics

The FWL flow rate $q_f(t)$ (m³/s) is governed by the pressure drop across the cake and filter medium, per Darcy's law [11]:

$$\frac{dq_f(t)}{dt} = \frac{1}{\tau_q} \cdot \left(\frac{\Delta P(t)}{R_{tot}(t)} - q_f(t)\right) \quad (7)$$

where $\tau_q$ is a flow response time constant (s), $\Delta P(t) = P_{atm} - P_v(t)$ is the driving pressure, and $R_{tot}(t) = R_c(t) + R_m$ is the total filtration resistance. $R_c(t)$ depends on $H(t)$, while $R_m$ is assumed constant. The model reflects that $q_f(t)$ responds rapidly to pressure and resistance changes, linking it to both $P_v(t)$ and $H(t)$ [14]. For instance, a sudden drop in $H(t)$ (due to cake discharge) reduces $R_c(t)$, increasing $q_f(t)$ until the cake rebuilds. Conversely, a vacuum increase (drop in $P_v(t)$) enhances $\Delta P(t)$, raising $q_f(t)$ momentarily. This equation completes the system dynamics by coupling the fluid and solid transport processes.

Finally, the nonlinear and coupled dynamic model is composed by (3)–(7). Table 1 lists the variables corresponding to the input and output currents of the system.

## 3.6 Steady-State Model

A steady-state model of the CD-filter can be derived from its nonlinear dynamic representation by nullifying all time derivatives in (3)–(7). This approach isolates the equilibrium operating conditions, which are fundamental for subsequent linearization and state- space analysis [15]. At steady-state, the system's variables satisfy a set of algebraic constraints representing torque equilibrium, vacuum balance, solids continuity, cake mass balance, and filtrate flow response. The resulting steady-state equations are:

$$\begin{cases} T_m^{ss} - k_d \cdot \omega^{ss} - k_c \cdot H^{ss} = 0 \\ Q_{air}^{in\,ss} - Q_{air}^{out\,ss} = 0 \\ F_{in}^{ss} \cdot C_{in}^{ss} - F_{out}^{ss} \cdot C_R^{ss} = 0 \end{cases}, \quad \begin{cases} \frac{1}{\rho_c \cdot A} \cdot C_R^{ss} \cdot Q_f^{ss} - \omega^{ss} \cdot H^{ss} = 0 \\ \frac{1}{R_{tot}} \cdot P_v^{ss} + Q_f^{ss} = -\frac{P_{atm}}{R_{tot}} \end{cases} \quad (8)$$

where all variables carrying the "ss" subscript denote steady-state values.

**Table 1. Input and output variables of the CD-filter system.**

| Variables | Symbol | Category |
|---|---|---|
| RWL inflow rate to the vat (m³/s) | $f_{in}(t)$ | |
| Solids concentration in incoming RWL (kg/m³) | $C_{in}(t)$ | Input |
| Air/vapor input flow into system (m³/s) | $q_{air}^{in}(t)$ | |
| FWL outflow rate through the filter media (m³/s) | $q_f(t)$ | |
| Solids concentration in the inflowing slurry (kg/m³) | $C_R(t)$ | Output |
| Air/vapor output flow from the system (m³/s) | $q_{air}^{out}(t)$ | |

To solve this system, the vector of unknowns is defined as: $\mathbf{q} = [T_m^{ss}\ Q_{air}^{in\,ss}\ F_{in}^{ss}\ H^{ss}\ Q_f^{ss}]$, where $\mathbf{q} \in \{\mathbb{R}^5\}$, which represents a nonlinear algebraic system of dimension five. All remaining parameters and variables (such as $C_R^{ss}$, $C_{in}^{ss}$, $P_v^{ss}$, $F_{out}^{ss}$, and $\omega^{ss}$) are assumed to be known from process specifications, design data, or instrumentation.

The analytical solution of system (8) yields the steady-state operating point (OPs) expressions given by:

$$\begin{cases} T_m^{ss} = \omega^{ss} + \frac{k_c \cdot (P_{atm} - P_v^{ss})}{A \cdot \rho_c \cdot R_{tot} \cdot \omega^{ss}} \\ Q_{air}^{in\,ss} = Q_{air}^{out\,ss} \\ F_{in}^{ss} = \frac{F_{out}^{ss} \cdot C_R^{ss}}{C_{in}^{ss}} \end{cases} ; \quad \begin{cases} H^{ss} = \frac{C_R^{ss} \cdot (P_{atm} - P_v^{ss})}{A \cdot \rho_c \cdot R_{tot} \cdot \omega^{ss}} \\ Q_f^{ss} = \frac{(P_{atm} - P_v^{ss})}{R_{tot}} \end{cases} \quad (9)$$

These expressions form a consistent steady-state modeling framework, enabling the evaluation of OPs for performance monitoring, control-oriented linearization, and further simulation or optimization analyses.

### 3.7 State-Space Linear Model

Following the derivation of OPs from (3)–(7), the nonlinear dynamics of the CD-filter system are linearized using a first-order Taylor expansion around the equilibrium conditions. This yields a continuous-time state-space model in deviation variables (dvs), expressed as:

$$\begin{cases} \frac{d\mathbf{x}(t)}{dt} = \mathbf{A_{dv}} \cdot \mathbf{x}(t) + \mathbf{B_{dv}} \cdot \mathbf{u}(t) \\ \mathbf{y}(t) = \mathbf{C_{dv}} \cdot \mathbf{x}(t) + \mathbf{D_{dv}} \cdot \mathbf{u}(t) \end{cases} \quad (10)$$

Here, $\mathbf{x}(t) \in \{\mathbb{R}^5\}$, $\mathbf{u}(t) \in \{\mathbb{R}^6\}$, and $\mathbf{y}(t) \in \{\mathbb{R}^5\}$ represent the deviation-form state, input, and output vectors, respectively. All states are assumed directly measurable, such that $\mathbf{x}(t) = \mathbf{y}(t)$. The vector definitions are: $\mathbf{x}(t) = \mathbf{y}(t) = [\omega^{dv}(t), P_v^{dv}(t), C_R^{dv}(t), H^{dv}(t), q_f^{dv}(t)]^T$ and $\mathbf{u}(t) = [T_m^{dv}(t), q_{air}^{in\,dv}(t), q_{air\,t}^{out\,dv}(t), f_{in}^{dv}(t), C_{in}^{dv}(t), f_{out}^{dv}(t)]^T$.

The state-space matrices are structured as:

$$\mathbf{A_{dv}} = \begin{bmatrix} -K_{21} & 0 & 0 & -K_{31} & 0 \\ 0 & 0 & 0 & 0 & 0 \\ 0 & 0 & -K_{43} & 0 & 0 \\ -K_{34} & 0 & K_{14} & -K_{44} & K_{24} \\ 0 & -K_{15} & 0 & 0 & -K_{25} \end{bmatrix}, \mathbf{B_{dv}} = \begin{bmatrix} K_{11} & 0 & 0 & 0 & 0 & 0 \\ 0 & K_{12} & -K_{12} & 0 & 0 & 0 \\ 0 & 0 & 0 & K_{13} & K_{23} & -K_{33} \\ 0 & 0 & 0 & 0 & 0 & 0 \\ 0 & 0 & 0 & 0 & 0 & 0 \end{bmatrix}, \quad (11)$$

$$\mathbf{C_{dv}} = \mathbf{I}_5, \text{ and } \mathbf{D_{dv}} = \mathbf{0}_{5 \times 6}$$

where $\mathbf{I}_5$ and $\mathbf{0}_{5 \times 6}$ denote the 5×5 identity matrix and the 5x6 null matrix, respectively. The matrices $\mathbf{A_{dv}}$, $\mathbf{B_{dv}}$, $\mathbf{C_{dv}}$, and $\mathbf{D_{dv}}$ represent the state transition matrix, input influence matrix, output mapping matrix, and direct transmission matrix, respectively. These matrices are structured within the following dimensional spaces: $\{\mathbf{A_{dv}}, \mathbf{C_{dv}}\} \in \mathcal{M}_{4 \times 4}\{K\}$ and $\{\mathbf{B_{dv}}, \mathbf{D_{dv}}\} \in \mathcal{M}_{5 \times 6}\{K\}$. The state, input, and output, vectors are denoted by $\{\mathbf{x}(t), \mathbf{y}(t)\} \in \{\mathbb{R}^5\}$ and $\mathbf{u}(t) \in \{\mathbb{R}^6\}$.

The constants $K_{ij}$ ($\{i, j\} \in \{1, 2, 3, 4\}$) are defined based on physical parameters and steady-state conditions:

$$\begin{cases} K_{11} = \frac{1}{J} \\ K_{21} = \frac{k_d}{J}, K_{12} = \frac{R_g \cdot T}{V_g}, \\ K_{31} = \frac{k_c}{J} \end{cases} \begin{cases} K_{13} = \frac{C_{in}^{ss}}{V_{vat}} \\ K_{23} = \frac{F_{in}^{ss}}{V_{vat}} \\ K_{33} = \frac{C_R^{ss}}{V_{vat}} \\ K_{43} = \frac{F_{out}^{ss}}{V_{vat}} \end{cases} \begin{cases} K_{14} = \frac{Q_f^{ss}}{\rho_c \cdot A} \\ K_{24} = \frac{C_R^{ss}}{\rho_c \cdot A} \\ K_{34} = \frac{H^{ss}}{\rho_c \cdot A} \\ K_{44} = \frac{\omega^{ss}}{\rho_c \cdot A} \end{cases} \quad (12)$$

The deviation variables follow the general form: $z^{dv}(t) = z(t) - Z^{ss}$, where $Z^{ss}$ is the steady-state value. The validity of this model assumes small perturbations around equilibrium, ensuring linear approximation holds [15].

## 4. CD-FILTER EFFICIENCY CHARACTERIZATION

The efficiency of a CD-filter is quantitatively assessed by evaluating the system's capacity to retain suspended solids from the RWL and discharge clarified filtrate as FWL. This performance metric is captured through a solids-based mass balance that defines filtration efficiency $\eta(t)$ as the ratio of retained solids to incoming solids over time [11]. Mathematically, this is expressed as follows:

$$\eta(t) = 100 \cdot \left(1 - \frac{q_f(t) \cdot C_R(t)}{f_{in}(t) \cdot C_{in}(t)}\right) \quad (13)$$

The formulation assumes a well-mixed vat with negligible bypass, ensuring that all incoming solids contribute to cake formation.

Filtration efficiency $\eta(t)$ is a pivotal performance metric linking hydrodynamic filtration, solids transport, and cake formation. Its real-time evolution reflects the coordinated impact of vacuum pressure, disc speed, and flow dynamics. Deviations in $\eta(t)$ indicate operational faults—such as media fouling, cake compaction, or vacuum insufficiency—making it essential for monitoring, diagnostics, and predictive control in CD-filter optimization.

## 5. CONTROL SYSTEM DESCRIPTION

To achieve high-performance closed-loop regulation of the CD-filter system, two control strategies are formulated: (i) a decentralized multiloop architecture based on PI compensators, and (ii) an advanced centralized MPC framework. Both strategies are constructed over the linear state-space formulation (10), derived from a first-order Taylor linearization of the nonlinear dynamic model around the OPs.

The decentralized PI-based topology, depicted in Figure 2(a), comprises three feedback control

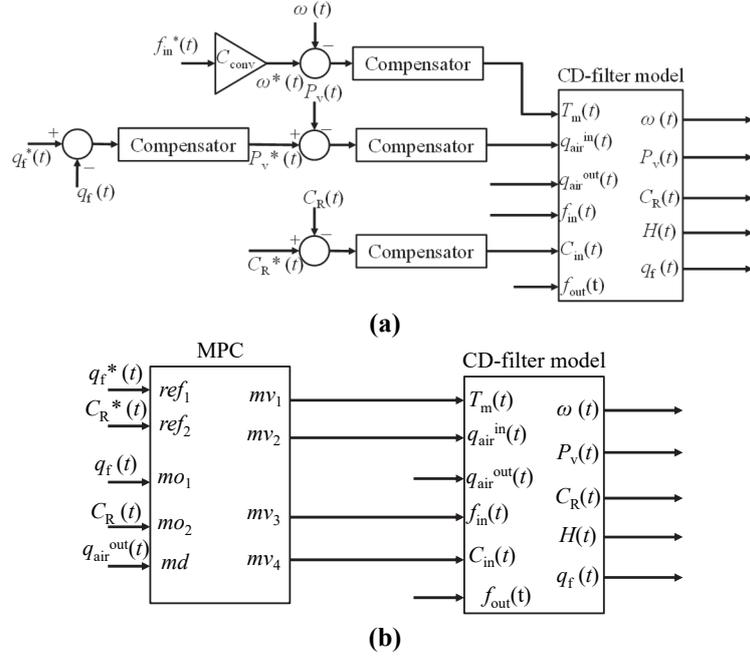

**(a)**

**(b)**

**Figure 2. Control structures proposed for the CD-filter operation. (a) Control law based on the implementation of PI compensators. (b) MPC control law.**

loops, each governing a key system output: angular velocity $\omega(t)$, solids in the slurry concentration $C_R(t)$, and FWL flow rate $q_f(t)$. This modular configuration enables distributed control over strongly coupled dynamics by leveraging physical interdependencies and steady-state relationships between variables. Recent studies have demonstrated that well-tuned multiloop PI architectures can offer high robustness and reliability in systems with moderate coupling and slow dynamics [16].

- Loop 1 – $\omega(t)$ regulation (via $T_m(t)$). The first loop regulates the disc's angular velocity $\omega(t)$, indirectly controlling the RWL feed flow rate $f_{in}(t)$. A steady-state gain conversion maps the inflow reference $f_{in}^*(t)$ to an equivalent angular velocity reference $\omega^*(t) = (F_{in}^{ss}/\omega^{ss}) \cdot f_{in}^*(t)$. The tracking error $e_\omega(t) = \omega^*(t) - \omega(t)$ is minimized using a PI controller, which generates the mechanical torque command $T_m(t)$ to adjust disc acceleration.

- Loop 2 – $C_R(t)$ regulation (via $C_{in}(t)$). This loop governs the solids concentration in the slurry tank by modulating the RWL inflow concentration $C_{in}(t)$. The error $e_{CR}(t) = C_R^*(t) - C_R(t)$ is processed by a second PI controller, whose output adjusts $C_{in}(t)$ in order to regulate the solids mass loading rate and maintain the desired concentration in the vat.

- Loop 3 – $q_f(t)$ regulation (via $q_{air}^{in}(t)$, $P_v(t)$) – cascade structure. The third loop adopts a cascade control configuration to regulate the FWL flow rate $q_f(t)$. The outer loop minimizes the error $e_{qf}(t) = q_f^*(t) - q_f(t)$ via a PI controller that computes the vacuum setpoint $P_v^*(t)$. The inner loop then regulates the actual vacuum $P_v(t)$ by manipulating the air/vapor inflow rate $q_{air}^{in}(t)$, using another PI controller. This dual-loop structure enhances control and compensates for the slower vacuum system dynamics.

In contrast, the model predictive control (MPC) architecture, shown in Figure 2(b), leverages the full multivariable linear state-space model to implement a predictive regulation scheme. At each sampling interval ($T_s$ = 0.1 s), the MPC solves a constrained quadratic optimization problem to compute the optimal control input sequence that minimizes a cost function comprising both output tracking errors and control effort penalties [17]. This structure explicitly accounts for coupling among $q_f(t)$ and $C_R(t)$, enabling coordinated control of the corresponding actuators: $T_m(t)$, $q_{air}^{in}(t)$, and $C_{in}(t)$.

Within the MPC structure, the dynamic coupling between inflow rate $f_{in}(t)$ and angular speed $\omega(t)$ via $T_m(t)$ is resolved internally through model-based decoupling. Optimal performance is achieved by prioritizing direct control of $q_f(t)$ and $C_R(t)$. In contrast, the PI architecture benefits from explicitly linking $f_{in}(t) \rightarrow \omega(t)$, improving coordination across subsystems. MPC further enforces hard input/output constraints (e.g., $T_m(t) \in [T_{m,min}, T_{m,max}]$, $C_R(t) \leq C_{Rmax}$), enabling safe and proactive disturbance rejection. This aligns with recent findings supporting MPC's superiority in tightly constrained, multivariable systems such as industrial filtration units [17].

## 6. SIMULATION RESULTS

In Figure 2(b), *mv* denotes manipulated variables, *mo* the measured outputs, *md* the measured disturbances, and *ref* the setpoints.

Closed-loop dynamic simulations were carried out in MATLAB-Simulink to evaluate the regulatory performance of two control schemes: decentralized multiloop PI and centralized MPC. The study targets the control of filtered liquor flow $q_f(t)$ and slurry concentration $C_R(t)$, which are actuated via vacuum inflow $q_{air}^{in}(t)$ and feed concentration $C_{in}(t)$, respectively. Simulations were conducted under time-varying reference trajectories representative of industrial transients, using parameters detailed in Table 2.

Figure 3 illustrates the closed-loop simulation results comparing the PI and MPC controllers. In Figure 3(a), the dynamic tracking of slurry vat solids concentration $C_R(t)$ is shown. The PI controller exhibits a rapid initial adjustment but with significant oscillations, leading to a high overshoot (~52 %) and an integral of squared error (ISE) of $1.72 \cdot 10^4$. In contrast, the MPC response achieves smoother transient behavior with reduced overshoot (~36%) and a lower ISE of $1.52 \cdot 10^4$. Both controllers converge quickly after the setpoint change, though MPC demonstrates superior damping and more accurate steady-state regulation. These results highlight MPC's enhanced capability in handling nonlinear coupling and improving robustness in multivariable CD-filter dynamics.

Also, from the Figure 3(a) the error distributions for $C_R(t)$ under both controllers are shown. The MPC error shows a sharper peak around zero with significantly reduced variance, yielding an 80.1% reduction in standard deviation compared to PI. This confirms MPC's superior precision and disturbance rejection, ensuring tighter regulation of slurry concentration.

Figure 3(b) presents the dynamic regulation of FWL $q_f(t)$. The PI controller achieves acceptable tracking but with a slight overshoot of ~1.97%, a settling time of 4 s, and an ISE of $1.10 \cdot 10^{-7}$. In contrast, the MPC response eliminates overshoot completely, reduces the settling time to 3 s, and maintains a comparable ISE of $1.11 \cdot 10^{-7}$. While both controllers deliver high-quality performance for $q_f(t)$, MPC demonstrates superior precision and faster stabilization, confirming its advantage in multivariable coordination and disturbance handling.

**Table 2. System parameters.**

| Parameter | Description | Value |
|---|---|---|
| $P_{atm}$ | Atmospheric pressure | 101.3 [kPa] |
| $J$ | Total shaft inertia | 3.5 [kg-m²] |
| $k_d$ | Viscous damping coefficient | 17.5 [Nm-s/rad] |
| $k_c$ | Cake resistance coefficient | 1,000 [Nm/m] |
| $t_q$ | Flow response time constant | 3 [s] |
| $R_g$ | Universal gas constant | 8.314 [J/mol-K] |
| $V_g$ | Receiver gas volume | 0.055 [m³] |
| $V_{vat}$ | Feed vat volume | 3 [m³] |
| $\rho_c$ | Cake bulk density | 1,050 [kg/m³] |
| $A$ | Filter area | 40 [m²] |
| $R_{tot}$ | Total filtration resistance | 125,000 [kPa-s/m²] |
| $T$ | Absolute temperature | 313 [K] |

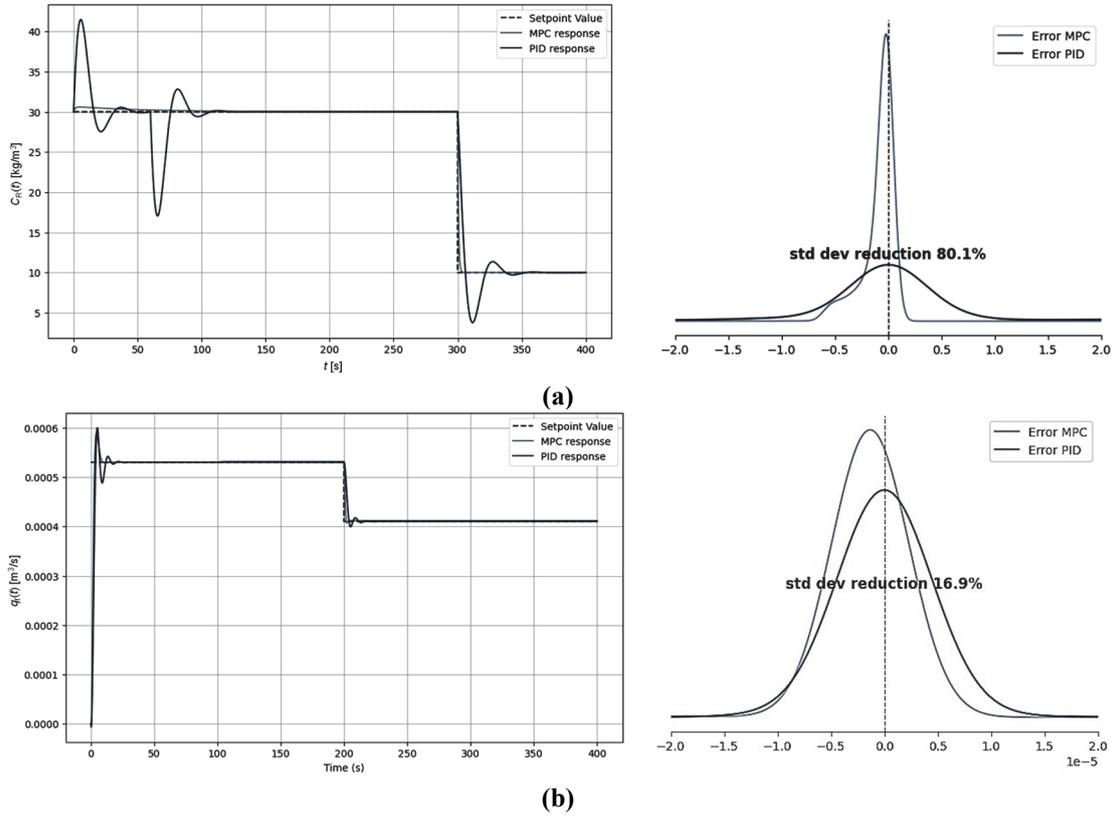

**Figure 3.** Closed-loop simulation results under transient dynamic for PI (dark blue) and MPC (light blue) controllers. Initial conditions of the system null. Step change in the reference of $q_f(t)$ $C_R(t)$ @ 200 s, and 300 s, respectively. (a) Solid concentration $C_R(t)$ tracking via $C_{in}(t)$ and the standard deviation information. (b) FWL flow $q_f(t)$ control via $q_{air}^{in}(t)$ and the standard deviation information.

Figure 3(b) also shows the error distributions of $q_f(t)$ under PI and MPC control. The MPC error distribution exhibits a narrower spread, yielding a 16.9% reduction in standard deviation relative to PI. This indicates improved precision and reduced variability in slurry concentration regulation, confirming MPC's tighter steady-state control.

A quantitative comparison of all performance metrics—ISE, overshoot, and settling time—is provided in Table 3. Across all indicators, the MPC outperforms the PI control strategy, particularly in concentration and vacuum loops where multivariable coupling is dominant.

Additionally, Figure 4 presents a 3D surface map of the filtration efficiency $\eta(t)$, calculated using (13). For constant values of $q_f(t)$ = 0.04 m³/s and $C_R(t)$ = 25 kg/m³, the efficiency increases nonlinearly with both $f_{in}(t)$ and $C_{in}(t)$. This surface highlights the sensitivity of $\eta(t)$ to feed rate and concentration, reinforcing the importance of coordinated control of input streams to achieve high-performance separation in CD-filter operation.

## 7. CONCLUSIONS

This work presented a nonlinear dynamic model of CD-filters, simplified to a single-disc equivalent for tractable control synthesis. The model captures the coupling between rotation, vacuum, slurry concentration, cake growth, and filtrate flow.

Comparative simulations showed that MPC outperforms PI by reducing ISE, overshoot, and settling time, especially in coupled variables such as vacuum and slurry concentration. Error analysis confirmed MPC's higher precision and lower variance.

The efficiency surface highlighted the need for coordinated control of feed flow and concentration. Overall, MPC proved superior in stability, robustness, and efficiency, making it a strong candidate for industrial CD-filter optimization.

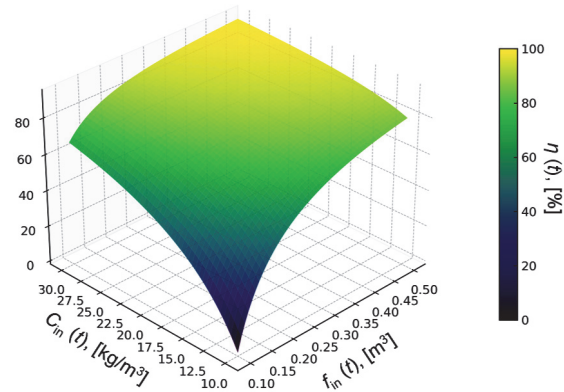

**Figure 4.** Filtration efficiency $\eta(t)$ [%] as a function of inlet flow rate $f_{in}(t)$ and inlet solids concentration $C_{in}(t)$, based on (13). With $q_f(t)$ = 0.04 m³/s and $C_R(t)$ = 25 kg/m³ held constant, the plot shows efficiency rising nonlinearly with both variables, emphasizing the need for coordinated control to enhance solids retention in CD-filter operation.

**Table 3.** Performance metrics summary.

| Variable | Controller | ISE | Overshoot [%] | Settling time [s] |
|---|---|---|---|---|
| $q_f(t)$ | PI | $1.10 \cdot 10^{-7}$ | 1.97 | 4.0 |
|  | MPC | $1.11 \cdot 10^{-7}$ | ~0 | 3.0 |
| $C_R(t)$ | PI | $1.72 \cdot 10^{4}$ | 52.0 | ~0 |
|  | MPC | $1.52 \cdot 10^{4}$ | 36.0 | ~0 |